\documentclass[pre,aps,floatfix,twocolumn,superscriptaddress,showpacs]{revtex4}
\usepackage{color}
\usepackage{graphicx}
\usepackage{amsmath}
\usepackage{amsthm}
\usepackage{amssymb}
\usepackage{latexsym}
\begin{document}
\date{\today}
\title{Force fluctuations in stretching a tethered polymer}
\author{Anoop Varghese}
\email{a.varghese@fz-juelich.de}
\affiliation{Institute of Complex Systems and Institute for Advanced
Simulation, Forschungszentrum J\"{u}lich, J\"{u}lich 52425, Germany}
\affiliation{The Institute of Mathematical Sciences, C.I.T. Campus,
Taramani, Chennai 600113, India}
\author{Satyavani Vemparala}
\email{vani@imsc.res.in}
\author{R. Rajesh}
\email{rrajesh@imsc.res.in}
\affiliation{The Institute of Mathematical Sciences, C.I.T. Campus,
Taramani, Chennai 600113, India}
\date{\today}
\begin{abstract}
The recently proposed fluctuation relation in unfolding forces [Phys.
Rev. E {\bf 84}, 060101(R) (2011)] is re-examined taking into account
the explicit time dependence of the force distribution. The stretching
of a tethered Rouse polymer is exactly solved and the ratio of the
probabilities of positive to negative forces is shown to be an
exponential in force. Extensive steered molecular dynamics simulations
of unfolding of deca alanine peptide confirm the form of fluctuation relation
proposed earlier, but with explicit correct time dependence of unfolding
forces taken into account. From exact calculations and simulations,
a linear dependence 
of the constant in the exponential of the fluctuation relation on
average unfolding forces and inverse temperature is proposed.   
\end{abstract}
\maketitle
\section{Introduction}

Fluctuation theorems provide a mechanism for characterizing 
fluctuations in non-equilibrium 
processes~\cite{bustamante2005,evans2002,jarzynski2008,sevick2008,jarzynski2011}.
These fluctuations become increasingly relevant as the system size
becomes smaller. Many biological systems are nano-sized and have inherently
non-equilibrium processes. Fluctuation theorems have been
realized in single molecule experiments such as dragging of a colloidal particle
in an optical trap~\cite{tft1,bead2004}, 
RNA unfolding experiments~\cite{liphardt2002,rnapulling}, 
and mechanical unfolding of proteins~\cite{shank2010,imparato2008}.

Non-equilibrium transient and steady states follow
transient~\cite{evans1994,evans2002,Zon2003,chakrabarti} (TFT) and steady
state (SSFT)~\cite{evans1993,gallavotti1995,searles1999,bonetto2001,zamponi2004,belushkin2011} 
fluctuation theorems respectively. In this paper,
we are concerned with TFT-like relation 
in unfolding forces of a tethered polymer. In TFT, the system is initially in an 
equilibrium state and fluctuations of quantities such as entropy, work, power flux 
and heat absorbed are measured over an arbitrary time interval~\cite{tft1,tft2,tft3,tft4,tft5,tft6}. 
For instance, the transient work fluctuation theorem~\cite{evans1994,kurchan2007} has the 
form $P(W)/P(-W)=e^{\beta W}$, where $P(W)$ is the probability of work
$W$ being done on the system. The Jarzynski's relation \cite{jarzynski1997}, 
which
relates the equilibrium free energy to non-equilibrium work, is a special form
of transient work fluctuation theorem, when the initial and final states are 
equilibrium states.

More recently, fluctuation theorems of non-traditional thermodynamic
variables like reaction coordinates~\cite{reaction,reaction1} and 
unfolding forces~\cite{ponmurugan}
have been studied. In Ref.~\cite{ponmurugan}, based
on constant velocity steered molecular dynamics (SMD) simulations of 
unfolding of contactin1 protein and deca alanine peptide, a fluctuation relation of the form 
\begin{equation}
\frac{P_v(+f)}{P_{v}(-f)} = \exp\left[\Gamma(T,v) f\right],
\label{eq:ponmurugan}
\end{equation}
was proposed, where, $v$ is the unfolding velocity and $f$ is the
unfolding force at constant temperature $T$. 
The constant $\Gamma(T,v)$ was observed to have the
scaling form 
\begin{equation}
\Gamma(T,v) \sim v^{\alpha} T^{-\delta}.
\label{eq:alphadefn}
\end{equation} 
For contactin1
protein $\alpha \approx 0.15$ and $\delta \approx 0.7$ and for deca
alanine $\alpha \approx 0.03$ and $\delta\approx 3.8$~\cite{ponmurugan}. 
However, analytical calculations for a Brownian particle in a harmonic
oscillator, moving at a constant velocity, show that though the form of
the fluctuation relation as proposed in Ref.~\cite{ponmurugan} is
retained, the exponents $\alpha$ and $\delta$ are equal to 
$1$~\cite{minh,sharma}. It is to be noted that 
while the SMD simulations~\cite{ponmurugan} were for a tethered molecule, 
the calculations~\cite{minh,sharma} are for a non-tethered particle. 
For a tethered molecule, the unfolding process is non-stationary and the fluctuation relation, if it
exists, should include an explicit time dependence. To address this issue, in this paper, we
solve exactly the time dependent force distribution $P_v(f,t)$ for a Rouse polymer and show that $P_v(f,t)$ is 
Gaussian and hence follows a fluctuation relation
\begin{equation}
\frac{P_v(+f,t)}{P_v(-f,t)}= \exp \left[ \Gamma \left( T,\langle
f(t)\rangle \right)  f  \right],
\label{eq:fluc_reln}
\end{equation}
with
\begin{equation}
\Gamma(T,\langle f(t)\rangle)=\frac{2\langle f(t)\rangle}{\alpha'T},
\label{eq:gamma}
\end{equation}
where $\langle f(t) \rangle$ is the time dependent average force 
and  $\alpha'$ is a system dependent constant. In case of
Rouse polymer, the time dependent average force is linear in
extension and hence unfolding velocity. We then perform
extensive SMD simulations of deca alanine in vacuum and show that, 
though the system has  non-linear force-extension
relation, the data for force distribution is still consistent with
Eq.~(\ref{eq:fluc_reln}). 

\section{Force distribution while stretching a rouse polymer}

In this section, we solve for the time dependent force distribution in a
tethered Rouse polymer. We closely follow the solution for the work
distribution derived in Ref.~\cite{dhar2005}.
Consider a one dimensional Gaussian chain consisting of $N+2$
particles.  
The particles are connected to each other by
harmonic springs such that the Hamiltonian of the system is 
\begin{equation}
H=\frac{k}{2}\sum_{i=1}^{N+1}\left(x_i-x_{i-1}\right)^2,
\end{equation}
where $x_i$ is the position of the $i^{\text{th}}$  particle, and $k$ is a constant. 
The first particle
is held fixed at the  origin and the last particle is pulled with a 
constant velocity $v$, {\it i.e.}, 
$x_0(t)=0$, and $x_{N+1}(t)=x_{N+1}(0)+vt$.
We assume Rouse dynamics, where the 
over damped Langevin equation for the chain is given by 
\begin{equation}
\frac{dx_i}{dt}=-\frac{k}{\gamma} (2 x_i - x_{i+1} - x_{i-1}) +\eta_i(t),
~i=1,2,\ldots N,
\label{eq0}
\end{equation}
where $\gamma$ is the friction coefficient and 
$\eta$ is white Gaussian noise with $\langle \eta_i(t)\rangle=0$ and $\langle
\eta_i(t)\eta_j(t') \rangle
=\frac{2}{\beta \gamma}\delta(t-t')$, where $\beta$ is the
inverse temperature. For convenience, we set $\gamma=1$. It can be
recovered in the later expressions by letting $k\rightarrow k/\gamma$
and $\beta \rightarrow \beta \gamma$.
Equation~(\ref{eq0}) may be written in matrix notation as
\begin{equation}
\frac{d{\bf x}}{dt}=-{\bf A} {\bf x}+{\bf h}(t)+{\bf \eta}(t).
\label{eq1}
\end{equation}
where ${\bf x}=(x_1,...,x_N)^T$, ${\bf \eta}=(\eta_1,...,\eta_N)^T$, and 
${\bf h}=(0,...,h_N)^T$, $h_N$ being $k\left[x_{N+1}(0)+v t\right]$.
{\bf A} is a tridiagonal symmetric matrix with non-zero entries
$A_{i,i}=2k$, $A_{i,i+1}=A_{i,i-1}=-k$. 

${\bf A}$ is diagonalized by an orthogonal transformation
${\bf O}^T  {\bf A O}={\bf \Lambda}$, where ${\bf O}^T$ = ${\bf O}^{-1}$ and 
${\bf \Lambda}$ is diagonal with $\Lambda_{mn}=\lambda_m\delta_{mn}$,
where $\lambda_m$'s are the eigenvalues of ${\bf A}$.
The eigenvalues $\lambda_m$ of ${\bf A}$, and the orthogonal matrix
${\bf O}$ are given by~\cite{mehtabook} 
\begin{eqnarray}
\lambda_m&=&2k\left[1-\cos\left(\frac{m\pi}{N+1}\right)\right],
\label{eq13}\\
O_{mn}&=&\sqrt{\frac{2}{N+1}}\sin\left(\frac{m n \pi}{N+1}\right).
\label{eq14}
\end{eqnarray}
Multiplying Eq.~(\ref{eq1}) with ${\bf O}^T$, we obtain
\begin{equation}
\frac{d\tilde {\bf x}}{dt}=-{\bf \Lambda} \tilde {\bf x}+\tilde {\bf
h}+ \tilde {\bf \eta},
\label{eq2}
\end{equation}
where $\tilde {\bf x}={\bf O}^T{\bf x}$, $\tilde {\bf h}={\bf O}^T {\bf h}$ and 
$\tilde {\bf \eta}={\bf O}^T {\bf \eta}$.
The general solution of Eq.~(\ref{eq2}) is 
\begin{equation}
\tilde {\bf x}(t)=e^{-{\bf \Lambda} t}\tilde {\bf x}(0)+\int_0^t dt'
e^{-{\bf \Lambda} (t-t')}\left[\tilde {\bf h}(t')+\tilde {\bf
\eta}(t')\right].
\label{eq3}
\end{equation}

The positions ${\bf x}(t)$ can be obtained from $\tilde {\bf x}(t)$
by ${\bf x} = {\bf O} \tilde {\bf x}$.  Since all the eigenvalues
of the matrix ${\bf A}$ are positive [see Eq.~(\ref{eq13})], 
the first term in Eq.~(\ref{eq3}) does not contribute in the limit of
large time and for convenience we set $\tilde
{\bf x} (0)={\bf 0}$.
Then, the position of the $N^{\text{th}}$ particle  is 
given by
\begin{equation}
x_N(t)=\sum_{m=1}^{N}O_{Nm}\int_0^tdt_1e^{-\lambda_m(t-t_1)}\left[\tilde
h_m(t_1)
+\tilde \eta_m(t_1)\right].
\label{eq5}
\end{equation}
The stretching force in the spring
connecting the $N^{\text{th}}$ and the $(N+1)^{\text{th}}$ particle
is given by 
$f(t)=k\left[x_{N+1}(t)-x_N(t)\right]=k\left[vt-x_N(t)\right]$. 
From Eq.~(\ref{eq5}) we see that $x_N$
is linear in the white noise $\eta$ and therefore its probability
distribution function will be a Gaussian. Likewise, $P_v(f,t$), the
distribution for force will be a Gaussian with $\langle f\rangle =
k\left[vt-\langle x_N(t)\rangle\right]$ and $\langle f^2
\rangle-\langle f\rangle^2 =  k^2[\langle x_N^2
\rangle-\langle x_N\rangle^2]$. 
We have to compute only the first two moments of
$x_N(t)$.

Averaging over noise in Eq.~(\ref{eq5}), and using 
$\tilde h_m= kvtO_{Nm}$, we obtain
\begin{equation}
\langle x_N(t)\rangle =
\sum_{m=1}^{N}O_{Nm}^2\frac{kv}{\lambda_m}
\left[t-\frac{1-e^{-\lambda_mt}}{\lambda_m}
\right],
\label{eq9}
\end{equation}
and
\begin{equation}
\langle x_N^2(t)\rangle-\langle
x_N(t)\rangle^2=\frac{1}{\beta}\sum_{m=1}^{N} O_{Nm}^2
\frac{1-e^{-2\lambda_mt}}{\lambda_m}.
\label{eq10}
\end{equation}

The results simplify in the limit of large time,
when the exponential terms in Eq.~(\ref{eq9}) and
Eq.~(\ref{eq10}) can be dropped. Then,
\begin{eqnarray}
\langle x_N(t)\rangle&=&vt\alpha_1-\frac{v}{k}\alpha_2,
\label{eq15}\\
\langle x_N^2(t)\rangle-\langle x_N(t)\rangle^2
&=&\frac{T}{k}\alpha_1,
\label{eq16}
\end{eqnarray}
where
\begin{eqnarray}
\alpha_1&=&k\sum_{m=1}^{N}\frac{O_{Nm}^2} {\lambda_m}\nonumber \\
&=&\frac{1}{N+1}\sum_{m=1}^{N}\frac{\sin^2\left(\frac{N\pi}{N+1}m\right)}{\left[1-\cos
\left(\frac{m\pi}{N+1}\right)\right]},
\label{eq17}
\end{eqnarray}
and
\begin{eqnarray}
\alpha_2&=&k^2\sum_{m=1}^{N}\frac{O_{Nm}^2}{\lambda_m^2}\nonumber \\
&=&\frac{1}{2\left(N+1\right)}\sum_{m=1}^{N}\frac{\sin^2\left(\frac{N\pi}{N+1}m\right)}{\left[1-\cos
\left(\frac{m\pi}{N+1}\right)\right]^2},
\label{eq18}
\end{eqnarray}
are constants which depend only on $N$.
Rewriting in terms of force, we obtain
\begin{eqnarray}
\langle f \rangle &=&kvt[1-\alpha_1]+v\alpha_2,
\label{eq19}\\
\sigma^2&=& \langle f^2 \rangle - \langle f \rangle^2 
=kT\alpha_1.
\label{eq20}
\end{eqnarray}
The force is then distributed as
\begin{equation}
P_v(f,t)=\frac{1}{\left(2\pi
kT\alpha_1\right)^{1/2}}\exp\left[-\frac{\left(f- \langle f \rangle \right)^2}
{2kT\alpha_1}\right],
\label{eq22}
\end{equation}
with $\langle f \rangle $ as in Eq.~(\ref{eq19}). Clearly,
\begin{equation}
\ln \left[ \frac{P_v(+f,t)}{P_v(-f,t)}\right]= \frac{2 f \langle f
\rangle}{ k T \alpha_1},
\label{eq:collapse}
\end{equation}
where the unfolding velocity $v$ is absorbed into $\langle f \rangle $
through Eq.~(\ref{eq19}). For the Rouse model considered here, 
$\langle f \rangle \propto v$,
and hence  Eq.~(\ref{eq:collapse}) has the
same form as the fluctuation relation Eq.~(\ref{eq:ponmurugan}) with
$\alpha=1$ and $\delta=1$, identical to the results obtained for a
Brownian particle in a harmonic oscillator~\cite{minh,sharma}.
However, if the dependence of $\langle f \rangle$ on $v$ is more
complicated, then the exponent in unfolding velocity, $\alpha$, may not be well defined.

\begin{figure}
\includegraphics[width=\columnwidth]{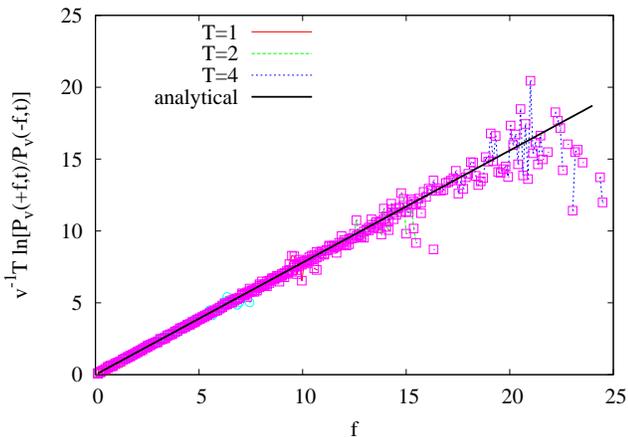}
\caption{(Color online)
The ratio $\ln[P_v(+f,t)/P_v(-f,t)]$ for
different temperatures and unfolding velocities for the Rouse model
collapse onto a single curve when
scaled as in Eq.~(\ref{eq:collapse}). The data are for $T=1, 2$ and
$4$.
The unfolding velocities are $1.0$ ($\Box$), $2.0$
($\circ$) and $4.0$ ($\triangle$).
The data points are averaged over $3\times10^6$ realizations. 
}
\label{fig2}
\end{figure}

We now do a numerical validation of the solution.
We numerically solve Rouse model by integrating the equations of
motion,
Eq~(\ref{eq0}), with initial condition $x_i(0)=0$ for all $i$. The force
$f$ is then obtained by $f=k(vt-x_N)$.
In Fig.~\ref{fig2} we show the data collapse
of $\ln[P_v(+f,t)/P_v(-f,t)]$ for various values of $v$ and $T$, when
scaled as in Eq.~(\ref{eq:collapse}) with $\langle f \rangle \sim v$. 
The agreement between the numerical and exact solutions is excellent.

Now, we would like to confirm whether Eq.~(\ref{eq:collapse}) holds for a more
realistic polymer. For contactin1 protein simulations done in 
Ref.~\cite{ponmurugan}, the prohibitive computational time due to
presence of large number of solvent molecules limited the number of
SMD runs from $3$--$30$. Obtaining reliable time dependent $P_v(f,t)$ from
such limited data set is not possible. On the other hand, deca alanine
in vacuum is a good test system that is computationally inexpensive. In the
next section, we describe set up and results of extensive SMD runs for
deca alanine.

\section{Steered MD simulations of deca alanine}

In this section, we describe the results of SMD 
simulations on deca alanine molecule, a prototypical system 
that has been used earlier for demonstrating
calculation of potential of mean force using Jarzynski's 
relation~\cite{park2003, park2004, ozer2012} and adaptive bias force 
methods~\cite{henin2004, henin2005}. 
Deca alanine molecule adopts a helical
conformation in vacuum and the SMD simulations have been performed by
fixing the C-terminal C$\alpha$ atom and unfolding the molecule by
pulling the N-terminal C$\alpha$ atom along the helical axis
with a constant velocity.
SMD simulations were performed at three different temperatures using
three different unfolding velocities (details
of simulation setup are given in Table~\ref{tabb2}). For each of the
parameter sets, 1000 SMD simulations were 
performed. Deca alanine was equilibrated for 10 ns
in vacuum in constant temperature-volume (NVT) 
conditions and from the last 5 ns of the run, 
the initial configurations for the SMD simulations were
generated by extracting 1000 snap shots.
All the
equilibrium and SMD simulations were performed by NAMD (version
2.7b3)~\cite{namd} using the CHARMM22 force field supplemented 
by CMAP corrections~\cite{cmap} for deca alanine. A cutoff of $12$\AA 
was used for van der Waals interactions and particle mesh
Ewald 
method was used to handle long range electrostatics interactions.
Langevin dynamics were used for temperature control 
and the box size was chosen to be large enough to accommodate the
stretched deca alanine molecule and to avoid interactions 
with the periodic images. A spring constant of 10 Kcal/mol/\AA$^2$ was
used for all the SMD runs. 
\begin{table}
\caption{\label{tabb2}
SMD simulation details of unfolding deca alanine for
different unfolding velocities $v$ and temperatures $T$ for spring
constant $K=10$ Kcal/mol/\AA$^2$.}
\begin{ruledtabular}
\begin{tabular}{ccc}
$v$    &    $T$     &  number of    \\
(\AA/ps)    &   (K)    &    SMD runs      \\
\hline
0.05        &   300    &   1000            \\
0.1        &   300    &   1000           \\
0.1        &   250    &   1000           \\
0.1        &   150    &   1000            \\
0.2        &   300    &   1000            \\
\end{tabular}
\end{ruledtabular}
\end{table}

\begin{figure}
\includegraphics[width=\columnwidth]{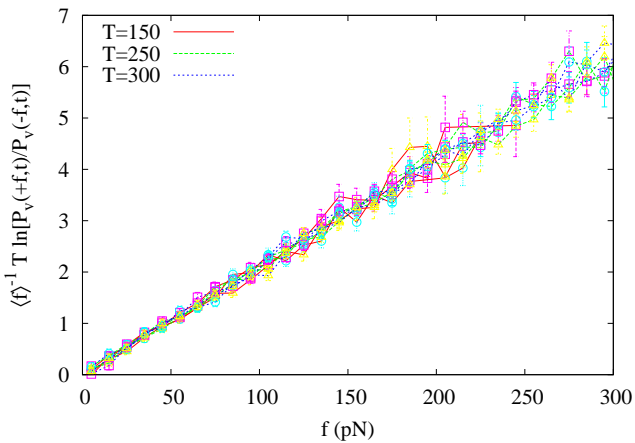}
\caption{\label{collapse_T}(Color online)
The ratio $\ln[P_v(+f,t)/P_v(-f,t)]$ for
different extensions and temperatures for deca alanine
collapse onto a single curve when
scaled as in Eq.~(\ref{eq:collapse}). The data are for $T=150, 250$ and
$300 K$.
The extensions are $2$\AA
($\Box$), $4$\AA ($\circ$) and $7$\AA ($\triangle$). All data are for unfolding
velocity $v=0.1$\AA/ps.
}
\end{figure}

\begin{figure}
\includegraphics[width=\columnwidth]{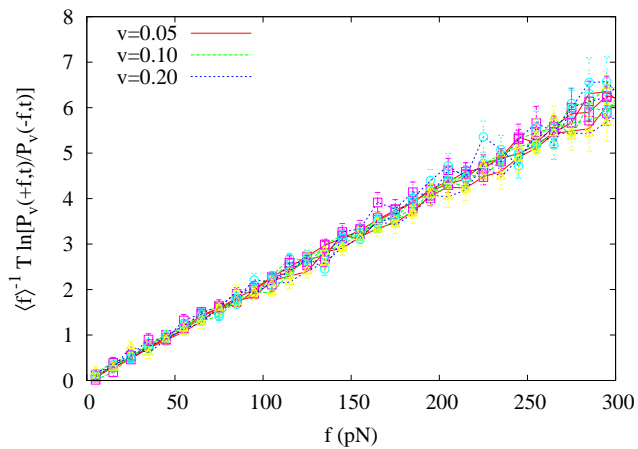}
\caption{\label{collapse_V}(Color online)
The ratio $\ln[P_v(+f,t)/P_v(-f,t)]$ for
different extensions and unfolding velocities  for deca alanine
collapse onto a single curve when
scaled as in Eq.~(\ref{eq:collapse}). The data are for $v=0.05,0.10$ and 
$0.20$\AA/ps. The extensions are $2$\AA
($\Box$), $4$\AA  ($\circ$) and $7$\AA ($\triangle$). All data are for 
temperature $T=300 K$.
}
\end{figure}

To construct $P_v(f,t)$, we consider all the unfolding forces
in a time window $(t-\Delta, t+\Delta)$ and average over $1000$ SMD
runs, where an optimum value of $\Delta$ is chosen such that we obtain
good statistics that are independent of $\Delta$. In
Fig.~\ref{collapse_T}, we show that the data for different
temperatures and times, for the same unfolding velocity, collapse on to
one curve when scaled as in Eq.~(\ref{eq:collapse}). Likewise, we find
good collapse for data for different unfolding velocities and same temperature
(see Fig.~\ref{collapse_V}).

\begin{figure}
\includegraphics[width=\columnwidth]{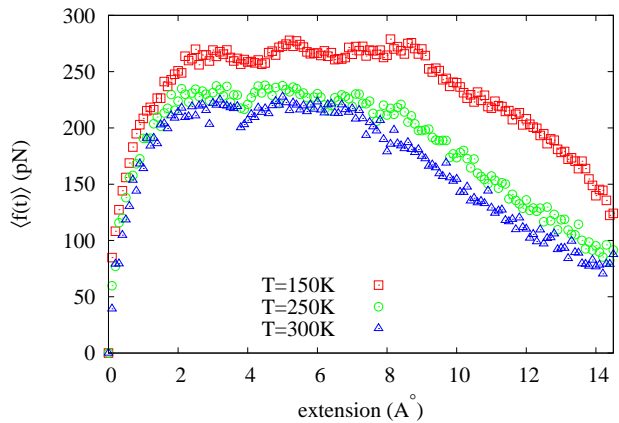}
\caption{\label{ave_force} (Color online) The average force as a function of
extension for different temperatures and fixed unfolding velocity 
$v=0.10$\AA/ps  for deca alanine .
}
\end{figure}

It is to be noted that for the collapse in Figs.~\ref{collapse_T} and
\ref{collapse_V}, we scaled the ratio of probabilities by the mean
force $\langle f \rangle$ rather than by $v$ as in Fig.~\ref{fig2}. 
The average force $\langle f\rangle$ for deca alanine is not a simple linear
function of the extension $v t$ (see Fig.~\ref{ave_force}). Therefore,
$\alpha$ [see Eq.~(\ref{eq:alphadefn}] is ill-defined for deca
alanine, though $\delta$ is seen to be $1$. From the exact
calculations and simulations, we expect that Eq.~(\ref{eq:gamma})
rather than Eq.~(\ref{eq:alphadefn}) will hold for the stretching of a
generic molecule.

\section{Conclusion}

In this paper, we re-examine the recently proposed fluctuation
relation, Eqs.~(\ref{eq:ponmurugan}) and (\ref{eq:alphadefn}),
in unfolding forces observed in the
SMD simulations of single molecules~\cite{ponmurugan}. Here, we
include the essential time dependence into the force distributions,
which was ignored in Ref.~\cite{ponmurugan}. First, we solved exactly
the time dependent force distribution for a tethered Rouse polymer that is
being unfolded at constant velocity. For this system, we obtain the
fluctuation relation 
Eq.~(\ref{eq:fluc_reln}) which has the same form as
Eq.~(\ref{eq:ponmurugan}) when the average unfolding force is proportional to
the unfolding velocity as is the case in the Rouse model. Second, using
extensive SMD simulations of deca alanine peptide in vacuum
for varying temperatures and unfolding velocities, we show that the data are
consistent with the fluctuation relation as in
Eq.~(\ref{eq:fluc_reln}) even though the average unfolding force is not
a simple function of unfolding velocity. The constant $\Gamma$
defined in Eq.~(\ref{eq:ponmurugan}) was proposed to be of the form
$v^{\alpha} T^{-\delta}$~\cite{ponmurugan}. Rather, we find
$\Gamma\propto \langle f \rangle T^{-1}$ as in Eq.~(\ref{eq:gamma}), 
where $ \langle f \rangle $
is a system dependent function of the unfolding velocity. It reduces to
the form $v T^{-1}$ for simple cases of a Brownian particle in a
harmonic potential ~\cite{minh,sharma} or the Rouse model considered
here.

If the time dependent force distribution is Gaussian, then the
fluctuation relation will have the form Eqs.~(\ref{eq:fluc_reln}) and
(\ref{eq:collapse}). 
In this paper, we showed that for a Rouse polymer the force
distribution is indeed Gaussian.
A priori, there is no
obvious reason to expect Gaussian distribution for a more realistic
polymer. However, for the prototypical deca alanine peptide studied here, 
the force distribution
appears to be Gaussian throughout the range of unfolding forces considered 
and also at various times along the unfolding trajectory making it
plausible that the force distribution is Gaussian
for an arbitrary molecule.

The proposed fluctuation relation in Ref ~\cite{ponmurugan} and its time dependent form 
in this paper, augment the list of fluctuation
relations (albeit in more conventional variables) in the literature. 
This may be realized in
the single molecule unfolding experiments.

\end{document}